\documentclass[aps,twocolumn,amsmath,letterpaper]{revtex4}
\usepackage{amssymb}
\usepackage{graphicx}

\newcommand{\Name}[1]{#1,}
\newcommand{\REVIEW}[4]{#1 {\bf #2}, #4 (#3)}

\allowdisplaybreaks[4]

\begin{document}
\title{Finite-temperature phase diagram of nonmagnetic impurities in high-temperature superconductors using a $d = 3$ $tJ$ model with quenched disorder}
\author{Michael Hinczewski$^{1}$ and A. Nihat Berker$^{1-3}$}

\affiliation{$^1$Feza G\"ursey Research Institute, T\"UBITAK -
Bosphorus University, \c{C}engelk\"oy 34680, Istanbul, Turkey,}
\affiliation{$^2$Department of Physics, Ko\c{c} University, Sar\i
yer 34450, Istanbul, Turkey,}
\affiliation{$^3$Department of Physics, Massachusetts Institute of
Technology, Cambridge, Massachusetts 02139, U.S.A.}

\begin{abstract}
  We study a quenched disordered $d=3$ $tJ$ Hamiltonian with static
  vacancies as a model of nonmagnetic impurities in high-$T_c$
  materials.  Using a renormalization-group approach, we calculate the
  evolution of the finite-temperature phase diagram with impurity
  concentration $p$, and find several features with close experimental
  parallels: away from half-filling we see the rapid destruction of a
  spin-singlet phase (analogous to the superconducting phase in
  cuprates) which is eliminated for $p \gtrsim 0.05$; in the same
  region for these dilute impurity concentrations we observe an
  enhancement of antiferromagnetism.  The antiferromagnetic phase near
  half-filling is robust against impurity addition, and disappears
  only for $p \gtrsim 0.40$.

\end{abstract}

\maketitle

The electronic properties and phase diagram of high-$T_c$ materials
are particularly sensititive to impurities---substitution of $3d$
transition elements (Zn, Ni, Co, Fe), or other metals (Al, Ga), for
the Cu atoms of the CuO$_2$ planes~\cite{Xiao}.  The interplay
between disorder, strong antiferromagnetic correlations in the
parent compound, and doped charge carriers, offers a window onto the
nature of both the superconducting phase and the normal state above
$T_c$. Doping by nonmagnetic ($S=0$) Zn ions provides one
representative example: the most pronounced effect is the rapid
destruction of the superconducting phase \cite{Xiao,Jayaram}; in
YBCO the transition temperature is reduced at a rate of $\sim
15$K/at.\% of impurities, so that it takes Zn concentrations of only
about $6\%$ to entirely eliminate superconductivity \cite{Jayaram}.
This is in contrast to the antiferromagnetic phase at half-filling,
which requires a far larger Zn concentration (about $40\%$ in
LSCO \cite{Vajk}) to completely suppress.  The effects in the
metallic region above $T_c$ are equally surprising: nuclear magnetic
resonance experiments have found that Zn atoms induce local magnetic
moments at nearest-neighbor Cu sites \cite{Mahajan}, and enhance
antiferromagnetic correlations for several lattice spacings around
the impurity \cite{Julien,Ouazi}.  In lightly hole-doped LSCO there
have been observations of an initial increase in the N\'eel
temperature with Zn addition, and even impurity-induced reappearance
of long-range antiferromagnetic order \cite{Hucker,Watanabe}.

In this work we model the effects of nonmagnetic impurities in
high-$T_c$ materials through a $d=3$ $tJ$ Hamiltonian with quenched
disorder in the form of static vacancies.  Through a
renormalization-group (RG) approach we obtain the evolution of the
global temperature versus chemical potential phase diagram with
disorder.  Our results capture in a single microscopic model some of
the major qualitative features of impurity-doping in real materials:
the rapid suppression of a spin-singlet phase, analogous to the
superconducting phase in cuprates; the gradual reduction of the
antiferromagnetic phase near half-filling; and the enhancement of
antiferromagnetism away from half-filling for small impurity
concentrations.

We consider the quenched disordered $tJ$ model on a $d$-dimensional
hypercubic lattice, $-\beta H = \sum_{\langle i j \rangle} \{-\beta
H_0(i,j)\} + \sum_i \mu^\text{imp}_i n_i$, where $-\beta H_0(i,j) = -t
\sum_{\sigma}(c^\dagger_{i\sigma}c_{j\sigma} +
c^\dagger_{j\sigma}c_{i\sigma})- J (\mathbf{S}_i\cdot\mathbf{S}_j
-n_i n_j/4) + \mu(n_i+n_j)$ is the standard $tJ$ model pair
Hamiltonian.  The static impurities at each site $i$ occur with
probability $p$ via $\mu_i^\text{imp} = -\infty$ and do not occur with
probability $1-p$ via $\mu_i^\text{imp} = 0$.

To formulate an RG transformation for this system, we use the $d=1$
Suzuki-Takano decimation \cite{SuzTak, TakSuz,
  FalicovBerker,FalicovBerkerT,Tomczak,TomRich1,TomRich2,HinczewskiBerker1,HinczewskiBerker2,KaplanBerker,Sariyer},
generalized to $d>1$ through the Migdal-Kadanoff method
\cite{Migdal,Kadanoff}.  This technique, adapted for quenched random
bond disorder, has recently elucidated the phase diagrams of the
quantum Heisenberg spin-glass in various spatial dimensions
\cite{KaplanBerker}.  In our case the rescaling for the $d=1$ system
(with sites $i=1,2,3,\ldots$) is:
\begin{align}
&\text{Tr}_\text{even} e^{-\beta
      H}\notag\\ &= \text{Tr}_\text{even} e^{\sum_{i}\left\{-\beta
        H_0(i,i+1)+\mu_i^\text{imp} n_i\right\}}\notag\\
&= \text{Tr}_{\text{even}} e^{\sum_{i}^{\text{even}}\left\{ -\beta
        H_0(i-1,i)+\mu_i^\text{imp}n_i-\beta H_0(i,i+1)
      \right\}+\sum_i^\text{odd} \mu_i^\text{imp}n_i}\notag\\
    &\simeq \left[\prod_{i}^{\text{even}}\text{Tr}_{i}e^{
        -\beta H_0(i-1,i)+\mu_i^\text{imp}n_i-\beta H_0(i,i+1) }\right]e^{\sum_i^\text{odd} \mu_i^\text{imp}n_i}\notag\\
     &=\left[\prod_{i}^{\text{even}}e^{-\beta^{\prime}
        H_0^{\prime}(i-1,i+1)}\right]e^{\sum_i^\text{odd} \mu_i^\text{imp}n_i}\notag\\
    &\simeq e^{\sum_{i}^{\text{even}}\left\{ -\beta ^{\prime
        }H_0^{\prime }(i-1,i+1)+\mu_{i-1}^\text{imp} n_{i-1}\right\} } = e^{-\beta ^{\prime }H^{\prime }},\label{eq:2}
\end{align}
where the traces and sums are over even- or odd-numbered sites $i$, and
$-\beta^\prime H^\prime$ is the renormalized Hamiltonian.
Anticommutation rules are correctly accounted for within segments of
three consecutive sites, at all successive length scales as the RG
transformation is iterated.

The algebraic content of the RG transformation is contained in the
second and third lines of Eq.~\eqref{eq:2}, yielding the renormalized
pair Hamiltonian $-\beta^\prime H_0^\prime (i^\prime,j^\prime)$
through the relation: $\exp(-\beta^\prime
H_0^\prime(i^\prime,j^\prime)) = \text{Tr}_{k} \exp( -\beta
H_0(i^\prime,k)+\mu_k^\text{imp}n_k-\beta H_0(k,j^\prime))$.  Under
the transformation the original system is mapped onto one with a more
general form of the pair Hamiltonian, $-\beta H_0(i,j) = -t_{ij}
\sum_{\sigma} (c^\dagger_{i\sigma}c_{j\sigma}+
c^\dagger_{j\sigma}c_{i\sigma})- J_{ij} \mathbf{S}_i\cdot\mathbf{S}_j
+ V_{ij} n_i n_j+ \mu_{ij} ( n_i +n_j) + \nu_{ij}(n_i - n_j)+G_{ij}$,
where the interaction constants $\mathbf{K}_{ij} \equiv (t_{ij},
J_{ij}, V_{ij}, \mu_{ij}, \nu_{ij})$ are nonuniform, and distributed
with a joint quenched probability distribution ${\cal
  P}(\mathbf{K}_{ij})$. This generalized form of the Hamiltonian
remains closed under further RG transformations.  Through the relation
above we can write the interaction constants
$\mathbf{K}^\prime_{i^\prime j^\prime}$ of the renormalized pair
Hamiltonian $-\beta^\prime H_0^\prime(i^\prime,j^\prime)$ as a
function of the interaction constants $\mathbf{K}_{i^\prime,k}$ and
$\mathbf{K}_{k,j^\prime}$ of two consecutive nearest-neighbor pairs in
the unrenormalized system, $\mathbf{K}^\prime_{i^\prime j^\prime} =
\linebreak[4] \mathbf{R}(\mathbf{K}_{i^\prime k},\mathbf{K}_{k
  j^\prime})$. This function $\mathbf{R}$ comes in two varieties,
depending on whether or not there is an impurity at site $k$, which we
shall denote as $\mathbf{R}_0$ and $\mathbf{R}_\text{imp}$
respectively.  Starting with a system with quenched probability
distribution ${\cal P}(\mathbf{K}_{ij})$, the distribution ${\cal
  P}^\prime(\mathbf{K}^\prime_{i^\prime j^\prime})$ of the
renormalized system is given by the decimation convolution
\cite{AndelmanBerker}: ${\cal P}^\prime(\mathbf{K}^\prime_{i^\prime
  j^\prime}) = \int d\mathbf{K}_{i^\prime k} d\mathbf{K}_{k
  j^\prime}\, {\cal P}(\mathbf{K}_{i^\prime k}) {\cal P}(\mathbf{K}_{k
  j^\prime}) [p\delta(\mathbf{K}^\prime_{i^\prime
    j^\prime}-\mathbf{R}_\text{imp}(\mathbf{K}_{i^\prime
    k},\mathbf{K}_{k j^\prime}))
  +(1-p)\delta(\mathbf{K}^\prime_{i^\prime
    j^\prime}-\mathbf{R}_0(\mathbf{K}_{i^\prime k},\mathbf{K}_{k
    j^\prime}))]$.  The initial condition for the RG flow is the
distribution corresponding to the original system, ${\cal P}_0
(\mathbf{K}_{ij}) = \delta(\mathbf{K}_{ij} - \mathbf{K}_0)$, where
$\mathbf{K}_0 = \{t,J,-J/4,\mu,0\}$.

\begin{figure}[t]
\centering \includegraphics*[width=0.48\textwidth]{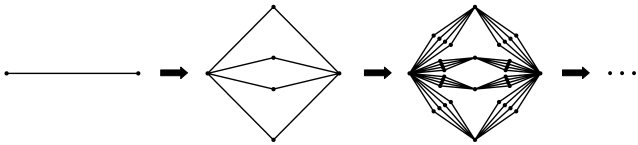}
\caption{Hierarchical lattice on which the $d=3$, $b=2$
  Migdal-Kadanoff recursion relations are exact.}\label{fig:1}
\end{figure}

The RG transformation is extended to $d>1$ through the Migdal-Kadanoff
\cite{Migdal,Kadanoff} procedure.  While approximate for hypercubic
lattices, the recursion relations generated by this procedure are
exact on hierarchical lattices
\cite{BerkerOstlund,GriffithsKaufman,Kaufman2}, and we shall use this
correspondence to describe the RG transformation for the case $d=3$,
with length rescaling factor $b=2$. The associated hierarchical
lattice is shown in Fig. 1.  Its construction proceeds by taking each
bond in the lattice, replacing it by the connected cluster of bonds in
the middle of Fig. 1, and repeating this step an infinite number of
times.  The RG transformation consists of reversing this construction
process, by taking every such cluster of bonds, decimating over the
degrees of freedom at the four inner sites of the cluster, yielding a
renormalized interaction between the two edge sites of the
cluster. Denoting these edge sites as $i^\prime$, $j^\prime$, and the
four inner sites as $k_1,\ldots,k_4$, this decimation can be expressed
as $\mathbf{K}_{i^\prime j^\prime} = \sum_{n=1}^4
\mathbf{R}(\mathbf{K}_{i^\prime k_n},\mathbf{K}_{k_n j^\prime})$. Just
as in the $d=1$ case, this decimation will give, after a single RG
transformation, a system with a nonuniform quenched distribution of
interaction constants.  We can calculate the quenched probability
distribution ${\cal P}^\prime(\mathbf{K}_{i^\prime j^\prime})$ of the
renormalized system through a series of pairwise convolutions,
consisting of the decimation convolution defined above for
interactions in series, and a ``bond-moving'' convolution for
interactions in parallel, using the function $\mathbf{R}_\text{bm}
(\mathbf{K}_A,\mathbf{K}_B) = \mathbf{K}_A + \mathbf{K}_B$.  In order
to numerically implement the convolution, the probability
distributions are represented by histograms, where each histogram is a
set of interaction constants $(t,J,V,\mu,\nu)$ and an associated
probability.  Since the number of histograms that constitute the
probability distribution increases rapidly with each RG iteration, a
binning procedure is used \cite{FalicovBerkerMcKay}.  Furthermore
since evaluation of the $\mathbf{R}$ functions is computationally
expensive, and most of the weight of the probability distributions is
carried by a fraction of the histograms, we have added an additional
step before the decimation convolution to increase efficiency: the
histograms with the 100 largest probabilities are left unchanged,
while the others are collapsed into a single histogram in a way that
preserves the average and standard deviation of the quenched
distribution. Thus we evaluate $10^4$ local decimations at each RG
transformation.

\begin{figure}
\centering
\includegraphics*[scale=1]{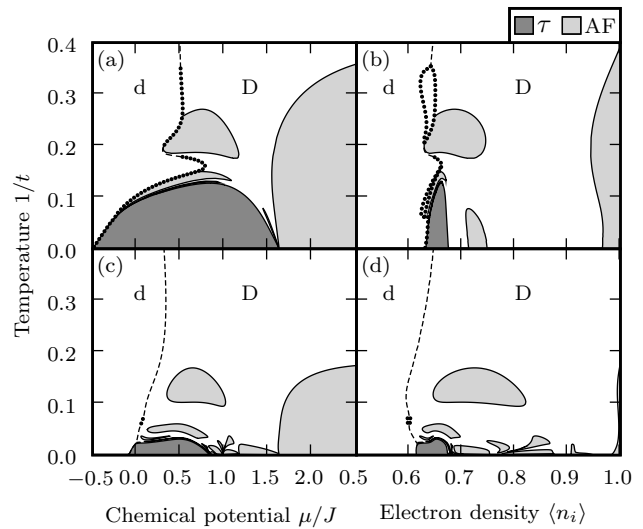}
\caption{Pure system ($p=0$) phase diagram of the isotropic $d=3$ $tJ$
  model \cite{FalicovBerker,FalicovBerkerT} for $J/t = 0.444$: (a) in
  terms of chemical potential $\mu/J$ vs.  temperature $1/t$; (b)
  electron density $\langle n_i \rangle$ vs.  temperature $1/t$.
  Panels (c) and (d) show the analogous phase diagrams for the
  uniaxially anisotropic case \cite{HinczewskiBerker2}, with
  $t_z/t_{xy} = 0.3$, $J_{z}/J_{xy} = 0.09$, $J_{xy}/t_{xy} = 0.444$.
  In both cases antiferromagnetic (AF), dense disordered (D), dilute
  disordered (d), and $\tau$ phases are seen.  The solid lines
  represent second-order phase transitions, while the dotted lines are
  first-order phase transitions (with the unmarked areas inside
  corresponding to coexistence regions of the two phases at either
  side).  Dashed lines are not phase transitions, but disorder lines
  between the dilute disordered and dense disordered
  phases.}\label{fig:2}
\end{figure}

All thermodynamic properties of the system, in particular the
finite-temperature phase diagram, can be determined from analyzing
the RG flows.  In the pure ($p=0$) case, the
transformation described above reduces to the recursion relations
derived for the $d=3$ $tJ$ model in earlier
studies \cite{FalicovBerker,FalicovBerkerT}, and yields the phase
diagram shown in Fig.~\ref{fig:2}(a,b) for $J/t = 0.444$.
Here we summarize the observed phases (for details,
consult \cite{FalicovBerker,FalicovBerkerT}): near half-filling
($\mu/J \to \infty$, $\langle n_i \rangle \to 1$), there is a
transition with decreasing temperature from a densely-filled
disordered phase (D) to long-range antiferromagnetic order (AF).
This AF phase persists away from half-filling down to $\mu/J \approx
1.6$, or 5\% hole doping.  For very large hole dopings ($\gtrsim
37\%$) we go over into a dilute disordered phase (d), with narrow
first-order coexistence regions between the d and D phases. At
intermediate hole dopings of 33-37\% a novel phase ($\tau$) is found
at low temperatures, flanked by an intricate lamellar structure of
AF islands.

\begin{figure}
\centering
\includegraphics*[scale=1]{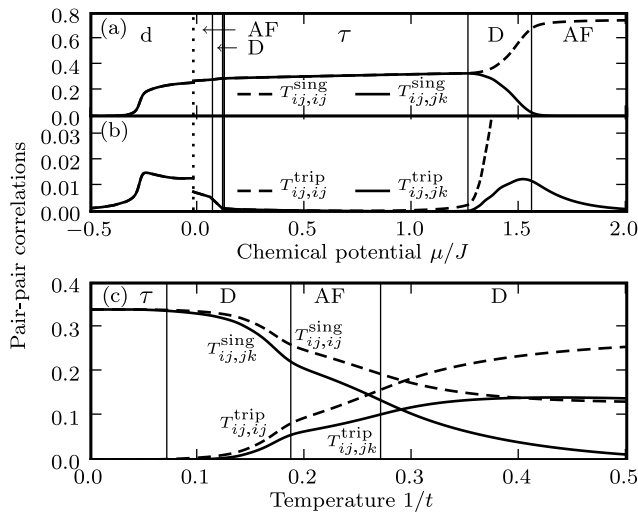}
\caption{On-site and nearest-neighbor singlet and triplet pair-pair
  correlations for the $d=3$ $tJ$ model, with $p=0$, $J/t = 0.444$.
  In (a) and (b) the correlations are plotted as a function of
  chemical potential $\mu/J$ at constant temperature $1/t = 0.10$.  In
  (c) they are plotted as a function of temperature $1/t$, at the
  constant electron density $\langle n_i \rangle = 0.67$.  The
  corresponding phases are indicated near the top of each plot, with
  solid and dotted vertical lines marking second-order and first-order
  phase boundaries respectively.}\label{fig:3}
\end{figure}

\begin{figure}
\centering
\includegraphics*[scale=1]{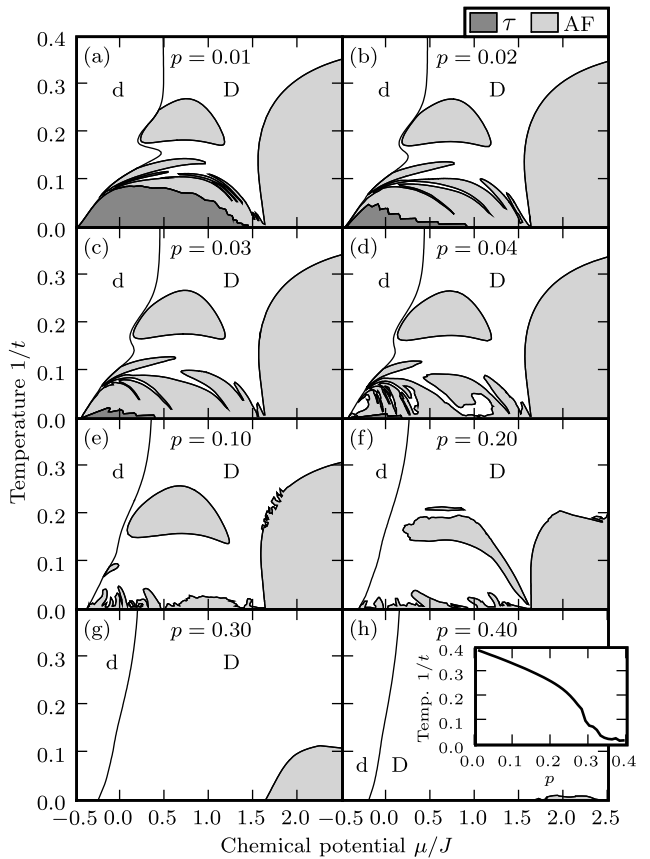}
\caption{Calculated phase diagrams of the $d=3$ \: $tJ$ model, with
$J/t = 0.444$, for various values of the impurity
concentration $p$, plotted in terms of temperature $1/t$ vs.
chemical potential $\mu/J$.  The phases depicted in
  the figures are: dilute disordered (d), dense disordered (D),
  antiferromagnetic (AF), and $\tau$.  The inset shows AF transition
  temperatures for the near-half-filled system ($\mu/J = 100$) as a
  function of $p$.}\label{fig:4}
\end{figure}

The $\tau$ phase is characterized by the formation of nearest-neighbor
spin-singlet pairs, as can be understood from correlation functions
calculated using the RG flows.  Let us define a singlet pair-pair
correlation function $T^\text{sing}_{ij,kl} = \langle
{\Delta^\text{sing}_{ij}}^\dagger {\Delta^\text{sing}_{kl}} +
{\Delta^\text{sing}_{kl}}^\dagger {\Delta^\text{sing}_{ij}} \rangle$,
where ${\Delta^\text{sing}_{ij}} = \frac{1}{\sqrt{2}}(c_{i\downarrow}
c_{j\uparrow} - c_{i\uparrow} c_{j\downarrow})$, and the analogous
triplet correlation function $T^\text{trip}_{ij,kl}$ in terms of
${\Delta^\text{trip}_{ij}} = c_{i\uparrow} c_{j\uparrow} +
\frac{1}{\sqrt{2}}(c_{i\downarrow} c_{j\uparrow} + c_{i\uparrow}
c_{j\downarrow}) + c_{i\downarrow} c_{j\downarrow}$.  For clusters of
three consecutive sites $i$, $j$, $k$ in the lattice, Fig.~\ref{fig:3}
shows the on-site correlations $T^\text{sing}_{{ij},{ij}}$,
$T^\text{trip}_{{ij},{ij}}$, and nearest-neighbor correlations
$T^\text{sing}_{{ij},{jk}}$, $T^\text{trip}_{{ij},{jk}}$.  In
Fig.~\ref{fig:3}(a) and (b), we see a constant temperature slice at
$1/t = 0.10$ as $\mu/J$ is varied.  There is a broad region of
chemical potentials away from half-filling, centered at the $\tau$
phase, where both the on-site and nearest-neighbor singlet
correlations are strong, in contrast to the triplet correlations,
which are suppressed in the same region.  We see similar behavior in
Fig.~\ref{fig:3}(c), where the correlations are plotted as a function
of temperature $1/t$ at a constant electron density $\langle n_i
\rangle = 0.67$.  As we decrease the temperature, approaching the
transition into the $\tau$ phase, there is a significant increase in
the singlet correlations and rapid decay of the triplet correlations.
Spin-singlet liquids, i.e., the hole-doped resonating valence bond
(RVB) state, have featured prominently in theories of high-$T_c$
superconductivity (for a review see Ref. \cite{Anderson}).  As we
shall see below, the behavior of the $\tau$ phase under
impurity-doping is analogous to that of the superconducting phase in
high-$T_c$ materials.

Though in this study we focus on the isotropic $d=3$ model, there is
evidence that general features of the $p=0$ phase diagram discussed
above persist even in the case of spatial anisotropy, with uniform
interactions ($t_{xy}$, $J_{xy}$) along the $xy$ planes and weaker
interactions ($t_{z}$, $J_z$) along the $z$ direction.  Through a
similar RG approach, using the more complicated hierarchical lattice
associated with a uniaxially anisotropic cubic lattice~\cite{Erbas},
it was found in particular that the $\tau$ phase continues to exist in
roughly the same doping range even for weak interplanar coupling,
though as expected the transition temperatures steadily decrease as
the coupling is reduced~\cite{HinczewskiBerker2}.  A representative
phase diagram, with $t_z/t_{xy} = 0.3$, $J_{z}/J_{xy} = 0.09$,
$J_{xy}/t_{xy} = 0.444$, is shown in Fig.~\ref{fig:2}(c,d). Thus the
$\tau$ phase may be relevant even in the strongly anisotropic regime
important for high-$T_c$ materials, which are characterized by weakly
interacting CuO$_2$ planes.

In Fig.~\ref{fig:4} we show the evolution of our calculated phase
diagram with increasing impurity concentration $p$.  The $\tau$ phase
is rapidly suppressed for $p=0.01$ through $0.04$
[Fig.~\ref{fig:4}(a)-(d)], and is no longer present by $p=0.05$. The
rate at which the $\tau$ phase disappears is comparable to the
reduction of $T_c$ with nonmagnetic impurities in cuprates, where
typically concentrations $\approx 2-6\%$ (depending on dopant) are
enough to eliminate superconductivity \cite{Xiao,Jayaram}.  As the
area of the $\tau$ phase recedes for these small impurity
concentrations, the region it formerly occupied is replaced by a
complex lamellar structure of the AF phase.  We can understand this
enhancement of antiferromagnetism through an RVB-like picture of the
$\tau$ phase \cite{Martins}: in the pure case the nearest-neighbor
singlets resonate in all possible arrangements along the bonds; when
an impurity is added some of these arrangements are ``pruned'',
because the bonds adjacent to the impurity can no longer accommodate
singlets. This inhibition of singlet fluctuations leads to enhanced
antiferromagnetic correlations around the vacancy.  Such local AF
enhancement near dilute nonmagnetic impurities has been observed
through NMR and NQR studies on Zn-doped YBCO \cite{Julien,Itoh}, and
supported theoretically by finite-cluster studies of the $d=2$
Heisenberg \cite{Bulut} and $tJ$ \cite{Poilblanc,Odashima} models.
More dramatically, in lightly hole-doped
La$_{2-x}$Sr$_x$Cu$_{1-z}$Zn$_z$O$_4$ (with $x = 0.017$) the N\'eel
temperature actually increases with the addition of Zn up to $z =
0.05$, before turning downwards again at higher $z$ \cite{Hucker}. A
similar, though smaller, effect has been found even at larger hole
dopings of $x = 0.115$ and $0.13$, with the $T_N$ increasing up to $z
= 0.0075$ \cite{Watanabe}.  In the case of $x = 0.13$, there is even
no long-range antiferromagnetic order for the Cu spins in the Zn-free
compound; it appears for $z > 0.0025$.  This reappearance of
long-range AF order upon addition of impurities, at small hole-dopings
away from half-filling where it does not exist in the pure case, was
replicated in the $d=2$ $tJ$ model using a self-consistent
diagrammatic approach \cite{KircanVojta}, and in the $d=2$ Hubbard
model with the dynamical cluster approximation \cite{MaierJarrell}.
Thus the enhancement of the AF phase away from half-filling, which we
find at small impurity concentrations, is consistent with previous
experimental and theoretical indications.

On the other hand for larger concentrations of impurities, the
dilution of the spins in the lattice becomes the dominant effect, and
eventually all long-range magnetic order is destroyed in the system.
We see this in Fig.~\ref{fig:4}(e)-(h), showing phase diagrams for
$p=0.10$ through $0.40$, and in the inset which plots the AF
transition temperature as a function of $p$ near half-filling ($\mu/J
= 100$).  In contrast to the $\tau$ phase, the AF phase around
half-filling is robust against impurity addition, and only disappears
for $p \gtrsim 0.40$.  Qualitatively similar behavior has been seen in
the half-filled compound La$_2$Cu$_{1-z}$Zn$_z$O$_4$, where Zn
concentrations of $z \approx 0.4$ are required to reduce the N\'eel
temperature to zero \cite{Vajk}, much larger than those needed to
eliminate superconductivity in the hole-doped material.

To summarize, we have applied an RG approach to the quenched
disordered $d=3$ $tJ$ model, and found the evolution of the phase
diagram as a function of impurity concentration.  The spin-singlet
phase away from half-filling is quickly destroyed through the addition
of small quantities of static vacancies, while antiferromagnetism in
the same region is enhanced.  The antiferromagnetic phase near
half-filling is less sensitive to impurity addition, and completely
disappears only at larger concentrations.  These results all have
close parallels in experimental results from cuprates.  The RG method
described here for dealing with quenched disorder in the $tJ$
Hamiltonian could be generalized to more complex systems: for example
the disordered Hubbard model, where the double-occupation of sites is
allowed through a finite electron-electron repulsion.  The role of
electron correlations and disorder in this system has led to
interesting phase diagram predictions~\cite{HeidarianTrivedi,
  Fazileh,Song}, which could be further explored with RG techniques.

This research was supported by the Scientific and Technological Research
Council of Turkey (T\"UBITAK) and by the Academy of Sciences of
Turkey.


\begin{thebibliography}{0}

\bibitem{Xiao} \Name{G. Xiao, M.Z. Cieplak, J.Q. Xiao, and C.L. Chien}
  \REVIEW{Phys. Rev. B}{42}{1990}{8752}.
\bibitem{Jayaram} \Name{B. Jayaram, S.K. Agarwal, C.V. Narasimha Rao,
    and A.V. Narlikar} \REVIEW{Phys. Rev. B}{38}{1988}{2903}.
\bibitem{Vajk} \Name{O.P. Vajk, P.K. Mang, M. Greven, P.M. Gehring,
    and J.W. Lynn} \REVIEW{Science}{295}{2002}{1691}.
\bibitem{Mahajan} \Name{A.V. Mahajan, H. Alloul, G. Collin, and J.-F.
    Marucco} \REVIEW{Phys. Rev. Lett.}{72}{1994}{3100}.
\bibitem{Julien} \Name{M.-H. Julien, T. Feh\'er, M. Horvati\'c, C.
    Berthier, O.N. Bakharev, P. S\'egransan, G. Collin, and J.-F.
    Marucco} \REVIEW{Phys. Rev. Lett.}{84}{2000}{3422}.
\bibitem{Ouazi} \Name{S. Ouazi, J. Bobroff, H. Alloul, and W.A.
    MacFarlane} \REVIEW{Phys. Rev. B}{70}{2004}{104515}.
\bibitem{Hucker} \Name{M. H\"ucker, V. Kataev, J. Pommer, J. Harra\ss,
    A. Hosni, C. Pflitsch, R. Gross, and B. B\"uchner} \REVIEW{Phys.
    Rev. B}{59}{1999}{725}.
\bibitem{Watanabe} \Name{I. Watanabe, T. Adachi, K. Takahashi, S. Yairi,
  Y. Koike, and K. Nagamine} \REVIEW{Phys. Rev. B}{65}{2002}{180516}.
\bibitem{SuzTak} \Name{M. Suzuki and H. Takano} \REVIEW{Phys. Lett.
    A}{69}{1979}{426}.
\bibitem{TakSuz} \Name{H. Takano and M. Suzuki} \REVIEW{J. Stat.
    Phys.}{26}{1981}{635}.
\bibitem{FalicovBerker} \Name{A. Falicov and A.N. Berker}
  \REVIEW{Phys. Rev. B}{51}{1995}{12458}.
\bibitem{FalicovBerkerT} \Name{A. Falicov and A.N. Berker}
  \REVIEW{Turk. J. Phys.}{19}{1995}{127}.
\bibitem{Tomczak} \Name{P. Tomczak} \REVIEW{Phys. Rev. B}{53}{1996}{R500}.
\bibitem{TomRich1} \Name{P. Tomczak and J. Richter} \REVIEW{Phys. Rev. B}{54}{1996}{9004}.
\bibitem{TomRich2} \Name{P. Tomczak and J. Richter} \REVIEW{J. Phys. A}{36}{2003}{5399}.
\bibitem{HinczewskiBerker1} \Name{M. Hinczewski and A.N. Berker}
  \REVIEW{Eur. Phys. J. B}{48}{2005}{1}.
\bibitem{HinczewskiBerker2} \Name{M. Hinczewski and A.N. Berker}
  \REVIEW{Eur. Phys. J. B}{51}{2006}{461}.
\bibitem{KaplanBerker} C.N. Kaplan and A.N. Berker, Phys. Rev. Lett. {\bf 100}, 027204 (2008).
\bibitem{Sariyer} O.S. Sar\i yer, A.N. Berker, and M. Hinczewski,
  Phys. Rev. B {\bf 77}, 134413 (2008).
\bibitem{Migdal} \Name{A.A. Migdal} \REVIEW{Zh. Eksp. Teor.
    Fiz.}{69}{1975}{1457} [\REVIEW{Sov. Phys. JETP}{42}{1976}{743}].
\bibitem{Kadanoff} \Name{L.P. Kadanoff} \REVIEW{Ann. Phys.
    (N.Y.)}{100}{1976}{359}.
\bibitem{AndelmanBerker} \Name{D. Andelman and A.N. Berker}
  \REVIEW{Phys. Rev. B}{29}{1984}{2630}.
\bibitem{BerkerOstlund} \Name{A.N. Berker and S. Ostlund} \REVIEW{J.
    Phys. C}{12}{1979}{4961}.
\bibitem{GriffithsKaufman} \Name{R.B. Griffiths and M. Kaufman}
  \REVIEW{Phys. Rev. B}{26}{1982}{5022R}.
\bibitem{Kaufman2} \Name{M. Kaufman and R.B. Griffiths} \REVIEW{Phys.
    Rev. B}{30}{1984}{244}.
\bibitem{FalicovBerkerMcKay} \Name{A. Falicov, A.N. Berker, and S.R. McKay} \REVIEW{Phys. Rev. B}{51}{1995}{8266}.
\bibitem{Anderson} \Name{P.W. Anderson, P.A. Lee, M. Randeria, T.M. Rice, N. Trivedi, and F.C. Zhang}
  \REVIEW{J. Phys.: Condens. Matter}{16}{2004}{R755}.
\bibitem{Erbas} A. Erba\c{s}, A. Tuncer, B. Y\"{u}cesoy, and A.N. Berker, Phys. Rev. E {\bf 72}, 026129 (2005).
\bibitem{Martins} \Name{G.B. Martins, M. Laukamp, J. Riera, and
    E. Dagotto} \REVIEW{Phys. Rev. Lett.}{78}{1997}{3563}.
\bibitem{Itoh} \Name{Y. Itoh, T. Machi, C. Kasai, S. Adachi,
    N. Watanabe, N. Koshizuka, and M. Murakami} \REVIEW{Phys. Rev. B}{67}{2003}{64516}.
\bibitem{Bulut} \Name{N. Bulut, D. Hone, D.J. Scalapino, and E.Y. Loh}
  \REVIEW{Phys. Rev. Lett.}{62}{1989}{2192}.
\bibitem{Poilblanc} \Name{D. Poilblanc, D.J. Scalapino, and W. Hanke}
  \REVIEW{Phys. Rev. Lett.}{72}{1994}{884}.
\bibitem{Odashima} \Name{S. Odashima and H. Matsumoto}
  \REVIEW{Phys. Rev. B}{56}{1997}{126}.
\bibitem{KircanVojta} \Name{M. Kir\'can and M. Vojta}
  \REVIEW{Phys. Rev. B}{73}{2006}{14516}.
\bibitem{MaierJarrell} \Name{Th. A. Maier and M. Jarrell}
  \REVIEW{Phys. Rev. Lett.}{89}{2002}{77001}.
\bibitem{HeidarianTrivedi} D. Heidarian and N. Trivedi,
  Phys. Rev. Lett. {\bf 93}, 126401 (2004).
\bibitem{Fazileh} F. Fazileh, R.J. Gooding, W.A. Atkinson, and
  D.C. Johnston, Phys. Rev. Lett. {\bf 96}, 046410 (2006).
\bibitem{Song} Y. Song, R. Wortis, and W.A. Atkinson, Phys. Rev. B
  {\bf 77}, 054202 (2008).
\end{thebibliography}
\end{document}